\documentclass[seceq]{ptptex}

\usepackage{graphicx}
\notypesetlogo               % Logo of PTP

\newcommand{\nc}{\newcommand}		% new command
\newcommand{\renc}{\renewcommand}	% re-new command
\nc{\nuc}[2]	{$^{#1}${#2}} 		% nucleus
\nc{\vc}[1]	{\mbox{\boldmath $#1$}}	% boldmath(vector)
\nc{\bra}	{\langle}		% bra state	
\nc{\ket}	{\rangle}		% ket state
\nc{\wtil}	{\widetilde}		% wide-tilde
\nc{\del}       {\partial}
\nc{\mapleft}[1]{			% something under arrow
 \smash{\mathop{\,			%
  \hbox to 1.2cm{\rightarrowfill}\, }\limits_{#1}}}
\nc{\mydraft}	{
	\oddsidemargin  +01mm 
	\evensidemargin +01mm
	\textheight 230mm
	\textwidth 155mm}
\begin{document}
\markboth{T. Myo, K. Kat\=o and K. Ikeda}{   
Tensor Correlation in $^4$He and Its Effect on the Doublet Splitting in $^5$He
}

\title{
Tensor Correlation in $^4$He and Its Effect on the Doublet Splitting in $^5$He
}

\author{
Takayuki MYO,$^{1,}$\footnote{\noindent E-mail : myo@rcnp.osaka-u.ac.jp}
Kiyoshi KAT\=O$^2$ %\footnote{\noindent E-mail : kato@nucl.sci.hokudai.ac.jp} 
and
Kiyomi IKEDA$^3$ %\footnote{\noindent E-mail : k-ikeda@postman.riken.go.jp}
}

\inst{
$^1$Research Center for Nuclear Physics (RCNP), Osaka University, Ibaraki 567-0047, Japan\\
$^2$Division of Physics, Graduate School of Science, Hokkaido University, Sapporo 060-0810, Japan\\
$^3$The Institute of Physical and Chemical Research (RIKEN), Wako 351-0198, Japan.
}

\recdate{November 21, 2004}

\abst{

We investigate the role of tensor correlation on the structures
of \nuc{4}{He} and its effect on the doublet splitting in \nuc{5}{He}. 
We perform a configuration mixing calculation in the shell model type bases 
to represent the tensor correlation for $^4$He.
It is found that our model describes the characteristics of the tensor correlation, 
which is represented by an admixture of the $0s_{1/2}$ configuration 
with a spatially modified $0p_{1/2}$ orbit. 
For $^5$He, we solve a coupled OCM equation for an extended $^4$He+$n$ model, 
while taking into account the tensor correlation in the $^4$He cluster.
It is shown that the tensor correlation produces the Pauli blocking, in particular, for the $J^\pi={1/2}^-$ state,
and its effect causes about half of the $p$-wave doublet splitting in $^5$He. 
This indicates that the strength of the effective spin-orbit interaction should be reduced 
by about half from the conventional one.
We obtain a reliable $^4$He-$n$ interaction, including the tensor correlation,
which further improves the behavior of the $d$- and $f$-wave phase shifts in the $^4$He+$n$ system.

}

\maketitle\section{Introduction}

The tensor force is an important ingredient in the nuclear force
derived from the meson theory, and it plays a characteristic role in
the nuclear structure. Actually, we know that the contribution of
the tensor force to the binding energy in \nuc{4}{He} is 
of the same order of magnitude as that of the central force\cite{Ak86,Ka01}.
An important point of variational calculations is that the trial wave function
should consist of not only the simple component with the $(0s_{1/2})^4$
configuration of the shell model but also the $D$-state component.
The latter has the set of angular momenta $(J,L,S)=(0,2,2)$ obtained by
applying the tensor operator to the $(0s_{1/2})^4$ state.
However, in many nuclear models, such a $D$-state component is not taken into account explicitly,
and the dominant effect of the tensor force is renormalized into central and spin-orbit terms
of the $G$-matrix as an effective interaction\cite{Ak72}.
In recent calculations in real space employing a realistic interaction,
the tensor force has been treated explicitly and shown to play a crucially important role
in the contribution to the binding energy\cite{Ka01,No00}.
However, beyond the results that the $D$-state components increase by 15\% in $^4$He and
the expectation value of the tensor force reaches about 60 MeV,
there is little information about more detailed connections between tensor correlations and 
nuclear structures.
It is a very important problem to understand the effect of the tensor force
on the nuclear structure in a physically transparent manner
by explicitly describing the tensor correlation in the model space.

Recently, Toki, Sugimoto and one of the present authors (Ikeda) made progress forward obtaining 
a description of the tensor correlation in the model space. 
Considering that the dominant term of the tensor force is expressed by a one-pion exchange potential, 
they showed that the tensor correlation can be described as 
that which causes the charge-parity mixing of the single-nucleon orbit mediated by the pion field. 
More precisely, the creation or annihilation of a virtual pion, which causes 
coupling between a pseudo-scalar meson and a nucleon, brings about an excitation 
that changes the parity and induces a crossing over the major shell, for example, of a nucleon in a closed configuration. 
Furthermore, because of the isovector character of a pion, 
the charge mixing of a single particle orbit is required to realize charge exchange excitations. 
Their extension of the concept of a single particle orbit of a finite nucleus is 
natural for incorporating the tensor force based on the mean field approach. 
They made a further improvement of the variation under projections of the parity and charge number\cite{Su04}. 
This projection procedure is very important to obtain a realization of the characteristics 
of the strong tensor correlation. One important reason for this is that the tensor correlation due to 
the pion exchange force is realized by the collection of tensor components up to a very high momentum
caused by two-particle excitations beyond a major shell. 
They applied this improved framework of the charge-parity projected Hartree-Fock method (CPPHF) to the 
alpha particle and obtained a satisfactory amount of the tensor correlation energy.\cite{Su04}

The basic purpose of this paper is to understand the essential effects of the tensor correlation 
by treating the tensor force explicitly. 
For this purpose, \nuc{4}{He} and \nuc{5}{He} are good objects of study, due to the simpleness of their structures. 
Therefore, in this paper, we re-investigate the role of the tensor correlation
in the structure of \nuc{4}{He} and study its effect on \nuc{5}{He}, 
by taking into account the characteristic results studied with CPPHF.\cite{Su04} 
In the present study, we use the simpler and more conventional shell model approach,
which we now explain. 
Here we extend the doubly closed shell configuration $(0s)^4$ to $(0s)^4 + (0s)^2(0p)^2$, at least 
up to $2p$-$2h$ configurations. Through this mixture of $2p$-$2h$ configurations, 
the tensor force can contribute explicitly to the binding energy of \nuc{4}{He}. 
These extended shell-model calculations are performed in the sense of perturbation theory 
for the nearly degenerate case. 
This approach is simple, but is convenient for investigating physical characters of the tensor correlation. 
The studies of the CPPHF\cite{Su04,Ak04} and the approach using 
antisymmetrized molecular dynamics (AMD)\cite{Do05} have shown that 
the optimum length of the mixed $p_{1/2}$ orbital is much narrower than that for a simple harmonic oscillator basis,
and that a pair configuration of the $p_{1/2}$ orbital generates the $D$-state (with $[J,L,S]=[0,2,2]$) in the total system. 
Therefore, we treat the length parameters ($b_{0s_{1/2}}$, $b_{0p_1/2}$ and $b_{0p_{3/2}}$) 
of the single particle orbits in our wave function as variational parameters. 
Through these variational studies of \nuc{4}{He}, 
we can check the applicability of our model approach. 
We also attempt to use several kinds of effective interactions 
including the tensor part in order to understand the characteristics of the tensor correlation.

For \nuc{5}{He}, the effect of the tensor force was shown at an early stage in the study of nuclear physics
to have a dependence on the orbital states occupied by the last neutron.
Terasawa\cite{Te60} performed a configuration mixing calculation for the \nuc{4}{He} part 
that is similar to the present one, and he found that the internal excitation of \nuc{4}{He} 
causes Pauli blocking in \nuc{5}{He}, which produces an orbital-state dependence.
As a result, the state-dependent effect is responsible for more than half of the observed doublet 
splitting of $p$-wave states.
Nagata et al.\cite{Na59} also succeeded in reproducing almost all of the doublet splitting
by adopting the tensor force of a pion theoretical one
through the analysis with the coupled resonating group method (RGM).

Here, we re-analyze the doublet splitting of $p$-waves within an extended \nuc{4}{He}+$n$ model,
in which we explicitly take into account the tensor correlation in the \nuc{4}{He} cluster 
by mixing $2p$-$2h$ configurations of $(0s)^2(0p)^2$ with the $(0s)^4$ configuration.
The $(0s)^2(0p)^2$ configurations are suppressed due to the Pauli principle
between a valence neutron and $0p$ orbital neutrons in the \nuc{4}{He} cluster.
This Pauli-blocking effect is treated in the framework of the orthogonality condition model (OCM).
We obtain a set of coupled OCM equations of the extended \nuc{4}{He}+$n$ model,
since \nuc{4}{He} is represented by the configuration mixing.
Thus, we are able to investigate to what extent the tensor correlation of \nuc{4}{He} generates
the Pauli blocking and results in the doublet splitting of $p$-waves in \nuc{5}{He}.
It is shown that this $(0s)^2(0p)^2$ excitation is related to the pion-like $0^-$ coupling 
nature of the tensor correlation\cite{To02,Og04},
because the $0^-$ coupling causes a distinctly large admixture of the $(0p_{1/2})^2(0s_{1/2})^2$ configuration in \nuc{4}{He}
and leads to a particularly strong Pauli-blocking effect on the $p_{1/2}$ valence neutron in \nuc{5}{He}.
This point was not clarified in previous studies\cite{Te60,Na59}.
Finally, we present the new \nuc{4}{He}-$n$ interaction including the tensor correlation in \nuc{4}{He}
to study the behavior of the partial waves of the \nuc{4}{He}+$n$ scattering up to $d$- and $f$-waves.
We show that the tensor correlation plays an important role in accounting for \nuc{4}{He}+$n$ scattering phenomena.

In \S\ref{sec:4He}, we present the model, which is able to describe the tensor correlation, 
and investigate the structure of \nuc{4}{He}.
We also discuss the effective interaction including the tensor force used in the present model.
In \S\ref{sec:5He}, we investigate the \nuc{4}{He}-$n$ interaction that leads to the doublet splitting of \nuc{5}{He}
by solving the coupled equation of the extended \nuc{4}{He}+$n$ model.
A summary is given in \S\ref{sec:summary}.
\section{Tensor correlation in \nuc{4}{He}}\label{sec:4He}

We examine the physical aspects of the tensor correlation in the
structure of \nuc{4}{He}.  The \nuc{4}{He} nucleus is a good test case,
because an exact four-body calculation can be carried out. 
In this calculation, the amount of the dissolution of \nuc{4}{He} from the
$(0s_{1/2})^4$ configuration is estimated to be more than 10\% for
the $D$-state probability. \cite{Ak86,Ka01} The tensor force tends
to change the parity and to flip the spin of single particle
orbital states of a nucleon due to the vertex operator of
$(\boldmath{\sigma}\cdot\vc{r})$. Hence, in \nuc{4}{He}, two orbits
of $0s$ and $0p$ are expected to be coupled by the tensor force.
We, therefore, extend the description of \nuc{4}{He} from $(0s)^4$
to $(0s)^4 + (0s)^2(0p)^2$, at least up to $2p$-$2h$
excitations\cite{To02,Su04}. Thus, the tensor force can contribute
explicitly to the nuclear structure through the mixture of
$2p$-$2h$ configurations\cite{Te60}.

\subsection{Model space and variational equations}\label{sec:4He_model}

In this study, we use the harmonic oscillator wave function
(h.o.w.f.) for \nuc{4}{He}. In this approach, it is easy to understand
the physical mechanism of the tensor correlation. Each single
particle h.o.w.f. is described using different length parameters
$\{b_\alpha\}$, where $\alpha$ indicates $0s_{1/2}$, $0p_{1/2}$ and
$0p_{3/2}$, and we treat the parameters in the set $\{b_\alpha\}$ as variational ones
to conveniently include the higher shell effect caused by the tensor force, 
as shown in Fig.~\ref{fig:4He_ene}. This is a kind of extension 
of the shell model suggested by the results of \nuc{4}{He} 
given in Refs.~\citen{Su04} and \citen{Ak04}. 
Neutron-rich He isotopes were studied in a similar framework by Kohno et al.\cite{Ko78}.

The \nuc{4}{He} wave function with $(J^\pi,T)=(0^+,0)$ for spin and isospin
can be expressed by a superposition of the
following configurations $\Phi_i$:
\begin{eqnarray}
    \Psi(^{4}{\rm He})
&=& \sum_{i=1}^{6}\ a_i\: \Phi_i
    \label{eq:WF_4He}\ ,
    \qquad
    \renc{\arraystretch}{1.5}
    \begin{array}{rcl}
        \Phi_1
&=&     (0s_{1/2})^4_{00}\ ,
        \\
        \Phi_2
&=&     \left[(0s_{1/2})^2_{01},(0p_{1/2})^2_{01}\right]_{00}\ ,
        \\
        \Phi_3
&=&     \left[(0s_{1/2})^2_{10},(0p_{1/2})^2_{10}\right]_{00}\ ,
        \\
        \Phi_4
&=&     \left[(0s_{1/2})^2_{01},(0p_{3/2})^2_{01}\right]_{00}\ ,
    \\
        \Phi_5
&=&     \left[(0s_{1/2})^2_{10},(0p_{3/2})^2_{10}\right]_{00}\ ,
        \\
        \Phi_6
&=&     \left[(0s_{1/2})^2_{10}, \left[(0p_{1/2})(0p_{3/2})\right]_{10}\right]_{00}\ .
    \end{array}
\end{eqnarray}
The subscripts 00, 01, 10 or 11 represent $J$ and $T$, spin and isospin for the two-nucleon
pair and for the total system, respectively.  Here, in order to maintain 
the orthogonality of the configurations $\Phi_i$, we choose the same value 
for the length parameters of orbits that possess the same quantum numbers
in every configuration $\Phi_i$; for example,
$b_{0s_{1/2}}$ in $\Phi_1$ is equal to $b_{0s_{1/2}}$ in $\Phi_2$.

The Hamiltonian of \nuc{4}{He} is given by
\begin{eqnarray}
    H
&=& \sum_{i=1}^A{t_i} - T_G
+   \sum_{i<j}^A v_{ij}\ ,
    \label{eq:Ham}
\end{eqnarray}
with
\begin{eqnarray}
    v_{ij}
&=& v_{ij}^C + v_{ij}^{T} + v_{ij}^{LS} + v_{ij}^{Clmb}\ ,
\end{eqnarray}
where the mass number $A$ is 4.  The effective $NN$ interaction
$v_{ij}$ consists of central ($v^C_{ij}$), tensor ($v^T_{ij}$), LS
($v^{LS}_{ij}$) and Coulomb ($v_{ij}^{Clmb}$) terms.
In this analysis, we do not explicitly treat the short-range correlation 
caused by the strong short-range repulsion of the central force. 
We use several kinds of effective interactions 
to check the dependence of the tensor correlation on the employed interactions.
One of the effective interactions 
is constructed from a realistic interaction based on the $G$-matrix theory viewpoint.
The component of the tensor force with momentum that exceeds than 
the cut off momentum of the $Q$ space is considered to be renormalized into the effective central force.
Following to Ref.\citen{Ak04}, this cut off momentum $k_Q$ is taken 
to be more than twice the Fermi momentum, 
because the high momentum region ($k>k_Q$) caused by the tensor force is
considered to be coupled strongly with the short-range correlation.
This also indicates that in the region satisfying $r\le 0.8$ fm, where the short-range correlation dominates, 
the effective interactions are strongly screened so strongly that the bare tensor force in the very short-range region disappears. 
Detailed properties of the effective interactions are discussed in \S\ref{sec:e_int}.

The wave function $\Psi=\Psi(^{4}{\rm He})$ in Eq.~(\ref{eq:WF_4He}) 
is used as a variational function.  
The variation of the energy expectation value with respect to $\Psi$ is given by
\begin{eqnarray}
\delta\frac{\bra\Psi|H|\Psi\ket}{\bra\Psi|\Psi\ket}&=&0\ ,
\end{eqnarray}
which leads to the following equations:
\begin{eqnarray}
    \frac{\del \bra\Psi| H - E |\Psi \ket} {\del b_\alpha}
&=& 0\ ,\quad
    \frac{\del \bra\Psi| H - E |\Psi \ket} {\del a_i}
=   0\ .
	\label{eq:vari}
\end{eqnarray}
Here, $E$ is a Lagrange multiplier corresponding to the total
energy. The parameters $b_\alpha$ (explicitly, $b_{0s_{1/2}}$, $b_{0p_{1/2}}$ and $b_{0p_{3/2}}$) 
appear non-linearly in the energy expectation value and $\{a_i\}$ are the coefficients defined in
Eq.~(\ref{eq:WF_4He}). We solve these two kinds of variational
equations in the following two steps. 
First, fixing the values of all the parameters $b_\alpha$, we solve the linear equation for $\{a_i\}$ as 
an eigenvalue problem for $H$ with the basis $\{\Phi_i;i=1,\cdots,6\}$. 
Next, for the obtained eigenvalue $E$, which is a function of $\{b_\alpha\}$, 
we search for the minimum value of $E(\{b_\alpha\})$. 
In all the calculational procedures, the c.m.motion is treated by subtracting 
the expectation value of the operator $T_G$ in the Hamiltonian given in Eq.~(\ref{eq:Ham}).

\subsection{Characteristic features of the tensor correlation in \nuc{4}{He}}\label{sec:4He-2}

Here, we study the characteristic features of the tensor correlation with regard to the results for \nuc{4}{He} 
by solving the variational equations (\ref{eq:vari}).
Before investigating the properties of the effective interaction, in the studies, 
we would like to understand the changes of the \nuc{4}{He} structure 
from that given by conventional $(0s)^4$ description by adding the tensor force.
We use the Volkov No. 2\cite{Vo65} and G3RS \cite{Ta68,Fu80} for the central and LS terms 
of the effective $NN$ interaction, respectively, 
both of which have often been used in many cluster model studies assuming the $(0s)^4$ configuration of \nuc{4}{He}.
The Majorana parameter of the Volkov No. 2 and the strength of the G3RS are chosen as 0.6 and 900 MeV, respectively, 
in order to reproduce the energy difference between $p_{1/2}$-$p_{3/2}$ orbits. As a tensor term, 
we use the Furutani-Tamagaki (FT) tensor force\cite{Fu80,Fu79}, which was used in the analysis of the 
nucleus with mass $A=4$ to reproduce the scattering data of the \nuc{3}{He}+$p$ system \cite{Fu79}.

First, we check the calculational results for the variational length parameters $\{b_\alpha\}$
at the energy minimum point.
It is confirmed that the values for $b_{0p_{1/2}}$ and $b_{0p_{3/2}}$ 
are almost the same at the point of minimal energy.  
Hence, in the following calculations, we choose the same value for
$b_{0p_{1/2}}$ and $b_{0p_{3/2}}$, and we have two variational length
parameters, $b_{0s}$ for the $s$ orbit and $b_{0p}$ for the $p$ orbit.

\begin{figure}[t]
\centering
\includegraphics[width=9.5cm,clip]{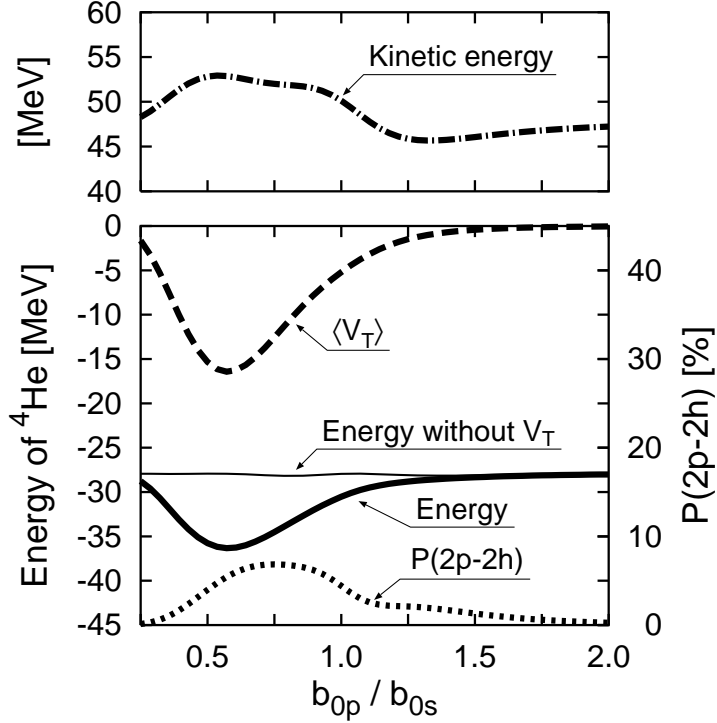}
\caption{The total energy of \nuc{4}{He} is plotted by the thick
solid curve as a function of the length parameter ratio,
$b_{0p}$/$b_{0s}$. 
The dashed and dash-dotted curves are the expectation values of the tensor force  
and the kinetic energy, respectively. 
The dotted curve is the probability of $2p$-$2h$ components $P(2p$-$2h)$ in the \nuc{4}{He} wave
function. The thin solid curve denotes the total energy of
\nuc{4}{He} calculated without the tensor force.}
\label{fig:4He_ene}
\end{figure}

The calculated results are shown in Fig.~\ref{fig:4He_ene} for the total energy, 
the expectation value of the tensor force $\bra V_T \ket$, and 
the mixing of the probability of the $2p$-$2h$ configuration states
as functions of the length parameter ratio, $b_{0p}/b_{0s}$.
Here, we fix $b_{0s}$ to 1.4 fm in order to reproduce the observed charge radius (1.68 fm).
The energy minimum is obtained at $b_{0p}\approx 0.6\ b_{0s}$. 
It is shown that $\bra V_T \ket$ has the largest contribution for values of the ratio $b_{0p}/b_{0s}$ 
near the energy minimum.
The $2p$-$2h$ configuration states are well mixed near the energy minimum.
The kinetic energies also realize their maximum values very near the energy minimal point.
However, in the range $0.3 < b_{0p}/b_{0s} < 1.5$, 
the amounts by which the kinetic energies increase and confined within a range of about 7 MeV,
which is almost half of the energy gain from $\bra V_T \ket$.
Then, the total energy can have an energy minimum pocket.
These results imply that the tensor force can be incorporated with
a small $b_{0p}$ value, which enables us to describe higher momentum components and
thereby higher shell effects of excited two-particle wave functions in $\{\Phi_i;i=2,3,\cdots,6\}$. 
It is noted that at the point $b_{0p}=b_{0s}$, that is, in the case of the standard shell model,
$\bra V_T \ket$ is small, about one-third of the minimal energy value of this calculation.
In Fig.~\ref{fig:4He_ene}, we also plot the total energies in the case that there is no tensor force,
and we see that the energy minimum disappears 
where the mixing of $2p$-$2h$ components caused by the central and LS forces is negligibly small, less than 2\% at most. 
In such a case, the $(0s)^4$ configuration of the ground state is considered to be approximated well.
These results for \nuc{4}{He} indicate that the tensor force plays
an important role in the dissolution from the double-closed
$(0s)^4$ configuration and leads to a tensor correlation through the mixing of $2p$-$2h$ configuration states.

Adoption of the tensor force in addition to the Volkov No. 2 central force
as the effective interaction leads to the overbinding of the total binding energy shown in Fig.~\ref{fig:4He_ene}.
This regards to be a double counting of the contribution of the tensor force, 
because the Volkov No. 2 central force is already renormalized by the tensor force.
However, a quantitative treatment of the total binding energy 
in which we solve the double counting problem is beyond the scope of the present model.
Here we postulate to adjust the total binding energy with the experimental one 
simply through the reduction of the strength of the attractive long-range part of the triplet-even ($^3E$) 
central term in the Volkov No. 2 by the factor $a(^3E)$,
since it is favorable to fit the binding energy for a quantitative understanding of the characteristic
features of the configuration mixing through detailed contributions of the mixed configurations.
The value of $a(^3E)$ is determined so as to fit the observed energy at the energy minimum, 
as shown in Table~\ref{tab:4He1}.
In this calculation, we also change $b_{0s}$ as a variational length parameter in addition to $b_{0p}$. 
The optimum values to obtain the energy minimum are ($b_{0s},~b_{0p}$)=(1.26,~0.77) fm.

\begin{table}[t]
\begin{center}
\caption{The properties of \nuc{4}{He} with the observed binding
energy (28.3 MeV) reproduced with two types of the strengths
of the tensor force. We list the contributions from each term in
the Hamiltonian (in units of MeV) and the matter radius ($R_m$, in units of fm) 
and the probabilities $P$ (in \%) of each configuration in Eq.~(\ref{eq:WF_4He}). 
The values $P(D)$ are the $D$-state probabilities.} \renc{\baselinestretch}{1.15} \label{tab:4He1}
\begin{tabular}{l|cc}\hline\hline
                            & $V_T \times$ 1.0 (original) & $V_T \times$ 1.5 (strengthen) \\ \hline
a($^3 E$)                   &     0.8379       &     0.6119       \\ \hline
$\bra T\ket$                &    65.21~\:      &   88.10~\:       \\ 
$\bra V_{C}\ket$     & $-$72.43~~~~     & $-$58.39~~~~     \\ 
$\bra V_{T}\ket$     & $-$22.62~~~~     & $-$60.65~~~~     \\ 
$\bra V_{LS}\ket$    &   0.66           &     1.67         \\ 
$\bra V_{Clmb}\ket$  &   0.89           &     0.96         \\ \hline
$R_m$                       &   1.34           &     1.22         \\ \hline
$P(2p$-$2h)$                &   7.24           &    16.81~\:      \\ 
~~~$\Phi_2$                 &   0.13           &     0.37         \\ 
~~~$\Phi_3$                 &   6.09           &    14.49~\:      \\ 
~~~$\Phi_4$                 &   0.07           &     0.19         \\ 
~~~$\Phi_5$                 &   0.75           &     1.67         \\ 
~~~$\Phi_6$                 &   0.20           &     0.09         \\ \hline
$P(D)$                      &   4.58           &    10.87~\:      \\ \hline
\end{tabular}
\end{center}
\end{table}

In Table~\ref{tab:4He1}, we list the properties of the \nuc{4}{He} wave functions
and obtain characteristics of the configuration mixing.
It is shown that the $2p$-$2h$ component of $\Phi_3$ in Eq.~(\ref{eq:WF_4He}) 
with an excitation to $(0p_{1/2})^2_{10}$ is strongly mixed. 
This result of the strong mixing of the $(0p_{1/2})^2_{10}$ excitation can be 
understood as follows: One particle in the $2p$-$2h$ components of $\Phi_3$ has
a dominant configuration caused by the $0^-$ coupling between the
orbits of $0s_{1/2}$ and $0p_{1/2}$. This $0^-$ coupling is
considered to exhibit the pion effect studied by Sugimoto et al.\cite{Su04} 
Furthermore, the two excited nucleons in $\Phi_3$
have the quantum numbers $(j,t)=(1,0)$ of spin and isospin,
respectively, which are the same as those of the deuteron. 
Therefore this two-particle coupling causing the $(0p_{1/2})^2_{10}$ mixing is
understood to be a deuteron-like correlation. That is the $(0s)_{10}^2$
component with quantum numbers $(L,S)=(0,1)$ for the orbital angular momentum and spin 
in the $(0s_{1/2})^4$ configuration of $\Phi_1$ in Eq.~(\ref{eq:WF_4He}) is strongly excited to
the $(0p_{1/2})_{10}^2$ component by the tensor force 
when changing the orbital angular momentum by 2 ($\Delta L=2$).
Here, $(0p_{1/2})_{10}^2$ has three components with $(L,S)$=(0,1), (2,1) and (1,0) 
in the $LS$ coupling scheme, and the middle part produces the $D$-state probability given in the last row of Table~\ref{tab:4He1}, 
in which the component of the set of the total orbital angular momentum $L_t=2$ and total spin $S_t=2$ 
is chosen in the total wave function of $\Psi(^{4}{\rm He})$ in Eq.~(\ref{eq:WF_4He}).
However, the obtained $D$-state probability is inconsistent with exact four-body calculations,
using realistic nuclear interactions which give about twice the present value.

In order to see effects of the tensor force on the $2p$-$2h$
components and the $D$-state probability, we perform the same
calculation with the matrix element of the tensor force strengthened by 50\%,
while the total energy is reproduced by changing $a(^3E)$ as shown in Table~\ref{tab:4He1}. 
The optimum values are found to be ($b_{0s}$, $b_{0p}$)=(1.14, 0.73) fm, which are slightly smaller than 
in the case of the tensor force with its original strength. 
The contribution of the tensor force, the amplitude of $\Phi_3$ and
the $D$-state probability increase by more than a factor 2 in comparison with the original ones. 
From these results, we confirm that the $\Phi_3$ component is strongly correlated with the tensor force.

\begin{figure}[t]
\centering
\includegraphics[width=9.5cm,clip]{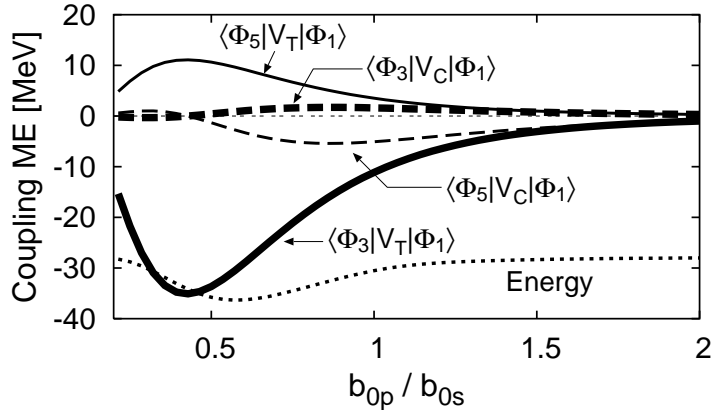}
\caption{
Coupling matrix elements of the central and tensor terms with respect to the $b_{0p}$/$b_{0s}$ ratio
in \nuc{4}{He}.
The configurations $[(0s_{1/2})^2_{10}(0p_{j})^2_{10}]_{00}$ are chosen for $2p$-$2h$ components, 
where $j$ is 1/2($\Phi_3$) and 3/2($\Phi_5$).
}
\label{fig:coupling}
\end{figure}

In Fig.~\ref{fig:coupling}, we display the $b_{0p}/b_{0s}$ dependence of the coupling matrix elements
between $\Phi_3$ and $\Phi_1$ and, for comparison, between $\Phi_5$ and $\Phi_1$, 
where the $2p$-$2h$ components have the same quantum numbers as the deuteron for the excited two-nucleon part.
Here we use the original strength of the tensor force and fix $b_{0s}$ to 1.26 fm.
It is found that the coupling matrix elements of the tensor force are enhanced for a narrow range of values of $b_{0p}$. 
In particular, the matrix elements between $\Phi_3$ and $\Phi_1$ are strongly increased for $b_{0p}/b_{0s}\sim$0.4.
It is also seen that the value of $b_{0p}/b_{0s}$ giving the point of minimal energy 
is somewhat larger than that for the point at the absolute maximum for the coupling matrix elements of the tensor force.
This results from an increase in the kinetic energy for small values of $b_{0p}$.

As shown in Table~\ref{tab:4He1}, wave functions give 
a matter radius smaller than the experimental value (1.48 fm) by 0.2 fm. 
This is due to the fact that $b_{0s}$ and $b_{0p}$ are narrow in comparison to the case
of the $(0s)^4$ single configuration without the tensor force.
Apparently, then, the tensor force tends to shrink a nucleus. 
This problem is considered to result from a problem with the effective
interaction employed here, and we discuss this in the next subsection.

To conclude this subsection, we can say that the present
model describes the characteristics of the tensor correlation in \nuc{4}{He} well. 
However, the saturation properties of \nuc{4}{He} cannot be studied quantitatively 
without detailed investigations of appropriate effective interactions. 
In order to make the studies of the tensor correlations more quantitative, 
we examine the effective interaction of the tensor part in more detail
in the following subsection.

\subsection{Effective interaction and tensor correlation}\label{sec:e_int}

Here we attempt to find an effective interaction that includes the tensor force and is suitable for the present model. 
Recently, from a similar viewpoint, Akaishi constructed the effective interaction 
based on the $G$-matrix theory using the realistic AV8' interaction \cite{Ak04} for \nuc{4}{He}.
In his prescription, the cut off momentum for the $Q$-space in the calculated $G$-matrices
is chosen as $k_Q=2.8$ fm$^{-1}$, which is larger than twice the Fermi momentum ($2k_F$), 
where $k_F$ is taken as 1.1 fm$^{-1}$ in \nuc{4}{He}. 
The reason why a very high cut off momentum
is adopted is that in the case of $D$-waves, momenta up to $\sim 2k_F$ are produced by the long-range part of 
the tensor correlations that appear in the realistic momentum distribution of nucleons in \nuc{4}{He}. 
In this new prescription, short-range correlations including very high momentum ($k>k_Q$) components of
tensor correlations are renormalized in the central term of the $G$-matrices. 
However, after screening the high momentum tensor components with $k>k_Q$, the tensor component in the $G$-matrices
survives strongly in the range around 0.8 fm with a width of 0.5 fm\cite{Ik04}, 
together with the one-pion tail in the truly long-range region. 
Furthermore, he discussed the fact that the renormalized central force having a one-pion tail exhibits 
a very weak density dependence\cite{Ik04}. 
Therefore, we use this interaction, called ``AK'', constructed by Akaishi, 
as a criterion for the effective interaction. 
We also refer to the so-called GPT interaction\cite{Go70}, which has Gaussian forms 
and is constructed to reproduce the two nucleon properties 
and to satisfy the charge radius of the doubly magic nuclei within the HF method.

\begin{figure}[t]
\centering
\includegraphics[width=13.0cm,clip]{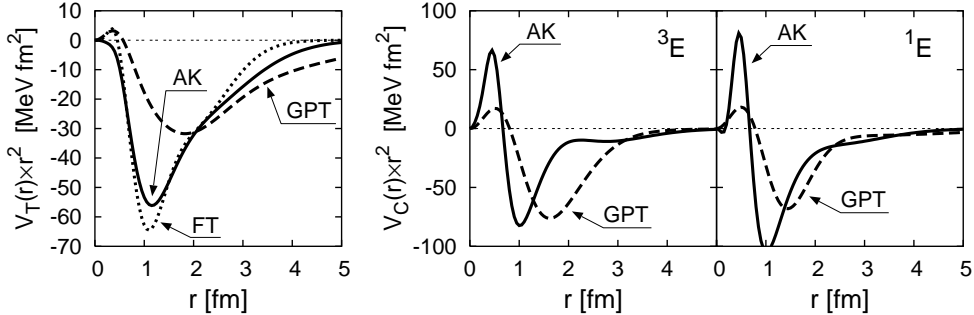}
\caption{Right: Three kinds of tensor forces for the triplet-even ($^3$E) state without the tensor operator part.
Left: Two kinds of central forces for $^3$E and singlet-even ($^1$E) states.}
\label{fig:int_CT}
\end{figure}

In order to demonstrate the properties of these interactions, 
we show in Fig.~\ref{fig:int_CT} radial forms of the tensor term without the tensor operator part and 
of the central term for two effective interactions, AK and GPT. 
For the tensor part, we also show the Furutani-Tamagaki force (FT) for comparison. 
It is found that AK and FT have similar shapes globally, but their tails behave slightly differently.
GPT has a shallow, long-range form. 
For the central part, AK has a spike in the short range and a small long tail. 
GPT has a mild short-range repulsion and a longer range than AK.
For AK, the $^3E$ component is smaller than the $^1E$ component in a volume integral,
because the renormalization from the tensor force into the central force is small in AK, 
in keeping with the character of the original AV8'. 
This feature is not seen for GPT. 
In GPT, renormalization of the tensor force is considered to be based on 
the conventional approach, which gives a strong $^3E$ component in the central force.

\begin{table}[t]
\caption{Properties of \nuc{4}{He} with three kinds of effective interactions.}
\label{tab:4He2}
\centering
\begin{tabular}{l|ccc}\hline\hline
                          &     AK       &   GPT        &    GA0       \\ \hline
($b_{0s}$,$b_{0p}$)~~[fm]~&~(1.12,0.65)~ &~(1.36,0.93)~ &~(1.26,0.75)~ \\ \hline
Energy~~[MeV]~            &  $-$19.3~~~~ &  $-$17.5~~~~ & $-$22.0~~~~  \\ \hline
$\bra V_T \ket$~~[MeV]    &  $-$32.5~~~~ &  $-$11.1~~~~ & $-$20.8~~~~  \\ \hline
$R_m$~~[fm]                & \,~1.20      & \,~1.46      &  \,~1.35     \\ \hline
$P(2p$-$2h)$~~[\%]        &   12.1~\,    &    8.3       &     8.0      \\ \hline
$P(\Phi_3)$~~~[\%]        &    8.3       &    6.8       &     5.6      \\ \hline
$P(D)$~~[\%]              &    6.3       &    5.1       &     4.3      \\ \hline
\end{tabular}
\end{table}

In Table~\ref{tab:4He2}, we show the results for \nuc{4}{He} obtained using AK and GPT in our model.  
Both interactions give small binding energies, because the higher-order correlations
are needed to reproduce the observed energy for these
interactions. AK gives a small radius with the present h.o.w.f.,
but a large tensor contribution. Since $k_Q > 2\ k_F$ in AK, we
can roughly estimate that $b_{0p} \sim b_{0s}/2 $ is allowed.
Contrastingly, GPT gives a good radius, but a small tensor
contribution. The difference between the two interactions mainly comes from
the strength of the tensor force. Since AK gives a small radius,
we also consider replacing the central part of AK by that of GPT. We
call this interaction ``GA0'', and its results are shown in Table~\ref{tab:4He2}. 
It is seen that the total energy is good and the radius has improved, but, it is still smaller than the experimental value.
For all three interactions used here, a large mixing of the $\Phi_3$ component in Eq.~(\ref{eq:WF_4He}) is found.
This implies that the characteristics of the tensor correlation do not depend strongly on the choice of the effective interaction.

It is noted that the tensor force of AK does not have a triplet-odd ($^3O$) component. 
This may contribute to the structure of \nuc{4}{He}. We study its effect using FT with
both even and odd components, instead of AK. The $^3O$ component
of the tensor force is found to contribute very little in the present model, 
about a few hundred keV in energy. 
From this result, we believe that this component does not play an essential role in the behavior considered in this study.

The above results show that neither AK nor GPT reproduces the saturation properties
of \nuc{4}{He} quantitatively.  It is necessary to improve the AK and
GPT interactions or to construct other appropriate interactions
that take into account the short-range and tensor correlations consistently in the present model space.

Here, we discuss the evaluation of the tensor correlation in our model 
in relation to the physical reason for strengthening the matrix elements of the tensor force,
here, e.g., by 50\%, as shown in Table~\ref{tab:4He1}. 
As seen in Fig.~\ref{fig:int_CT}, the tensor force multiplying $r^2$ 
not only has a large value in the pocket around $r\sim 1.0$ fm ($0.6<r<1.4$) 
from the short- to intermediate-range regions
but also has an appreciable value in the widely extended long-range region
$1.4<r<3.0$, which exceeds the average distance between two nucleons. 
However, our model wave function based on the so-called $V$-type basis 
is not necessarily sufficient to treat the tensor correlation from the short- to the long-range regions simultaneously,
and it is found that the present $V$-type basis favorably describes the tensor correlation 
from the deep pocket in the intermediate region with the narrow $0p$ orbit solution.
This problem is expected to be overcome by adding $T$-type basis wave functions, 
such as $[\phi_2(\vc{r},b_r),\phi_0(\vc{R},b_R)]_2$, 
with arbitrary h.o. length parameters ($b_r$, $b_R$), where $(\vc{r}, \vc{R})$ are the coordinates of the
relative and c.m.motion of the two nucleons, respectively, and the subscripts represent the orbital angular momenta. 
This basis is suitable for describing the behavior of the $D$-state 
in the wave function emerging from the tensor force, especially, in the short- and intermediate-range regions.
In the hybrid wave functions combined with $V$-type and $T$-type bases,
the $V$-type wave function expanded with several Gaussian bases\cite{Ao95}
is expected to be suitable for describing the long-range part of the tensor correlation.

To strengthen the matrix element of the tensor force by 50\% in the present $V$-type basis,
it is helpful to consider the relation between $V$-type and $T$-type bases.
They can overlap, and the $(0p)^2$ component in the $V$-type basis can be expanded 
in terms of the $T$-type basis as
\begin{eqnarray}
    (0p)^2_{L=2}
&=& \frac{1}{\sqrt{2}}\biggl( [\phi_2(\vc{r},b_r'),\phi_0(\vc{R},b_R')]_2
+ [\phi_0(\vc{r},b_r'),\phi_2(\vc{R},b_R')]_2 \biggr),
    \label{eq:VT}
\end{eqnarray}
where the spin part is ignored for simplicity and the h.o. length parameters are related to 
$b_{0p}$ of the $0p$-orbit by $b_r'=\sqrt{2}\cdot b_{0p}$ and $b_R'=(1/\sqrt{2})\cdot b_{0p}$.
This relation shows that the first term on the right-hand side of
Eq.~(\ref{eq:VT}) of the $T$-type basis can be reduced by a factor of $1/\sqrt{2}$. 
Since we restrict the model wave function to that of the shell model of the $V$-type basis in this study, 
we postulate the strengthening of the matrix elements of the tensor force by an appreciable amount, 
which is assumed to be $50$\%. Justification of the present argument is given 
by extending the analysis of \nuc{4}{He} using the hybrid wave functions
of the $V$-type and of $T$-type.

\section{Tensor correlation in \nuc{5}{He}: doublet splitting}\label{sec:5He}

In the nucleus, a large doublet (LS) splitting is observed, 
and this can be explained by a strong spin-orbit (LS) force. 
For $p$-shell nuclei, the doublet splitting of $p$-wave resonances
($3/2^-$ and $1/2^-$) in \nuc{5}{He} has been studied 
as a way of finding the strength of the LS force, because \nuc{5}{He}
is the lightest nucleus in which the doublet splitting is
observable and is considered to be the simplest nucleus to have one
nucleon above the double-closed $(0s)^4$ core. However, as shown
in the previous section, the tensor force produces a mixture of
a large amount of the $D$-state component with the $(0s)^4$
configuration in \nuc{4}{He}.

\subsection{A tensor-correlated \nuc{4}{He}-cluster plus neutron model}

We analyze the $p$-wave resonances of \nuc{5}{He} with an extended
\nuc{4}{He}+$n$ model, where we explicitly take into account the
tensor correlation in the \nuc{4}{He} cluster by mixing $2p$-$2h$
components with the $(0s)^4$ configuration. 
In this analysis, we are chiefly interested in the coupling between the internal
structure of the \nuc{4}{He} cluster and the motion of the $p$-wave valence neutron. 
In particular, the $2p$-2$h$ (tensor) correlation of \nuc{4}{He} is considered to be 
suppressed due to the Pauli principle with the valence neutron, and therefore the
\nuc{4}{He} cluster in \nuc{5}{He} loses energy as compared 
to the isolated case. This is the so-called Pauli-blocking effect,
which is similar to the case induced by the pairing correlation in
the \nuc{9}{Li}+$n$ system \cite{Ka99,My02}.

From the results for \nuc{4}{He}, we learn that $\Phi_3$ including
the $0p_{1/2}$ orbit in Eq.~(\ref{eq:WF_4He}) is strongly mixed.
Therefore, the large $0p_{1/2}$ mixing in \nuc{4}{He} results in a stronger
Pauli-blocking effect for the $1/2^-$ state of \nuc{5}{He} than
for the $3/2^-$ state.
For this reason, we expect that the $1/2^-$ state of
\nuc{5}{He} experiences a larger energy loss for the \nuc{4}{He} cluster
than does the $3/2^-$ state and as a result leads to doublet
splitting. This has also been investigated qualitatively in previous studies\cite{Te60,Na59}, 
where about a half of the doublet splitting comes from the tensor correlation.

In order to evaluate the Pauli-blocking effect in \nuc{5}{He}
correctly, it is necessary to carry out the configuration mixing
in the \nuc{4}{He} cluster and the antisymmetrization between
nucleons in the \nuc{4}{He} cluster and the valence neutron $n$. 
In order to accomplish this, we start by adopting the following wave function of the resonating group method (RGM) with
six \nuc{4}{He}-cluster configurations:
\begin{eqnarray}
    \Psi^J(^5{\rm He})
&=& \sum_{i=1}^6 {\cal A}\left\{ [\Phi_i(^4{\rm He}),\chi_i^J(r)]^J \right\}\ .
    \label{eq:WF_5He}
\end{eqnarray}
Here ${\cal A}$, $r$, $\{\chi^J_i(r)\}$ and $J$ are
an anti-symmetrizer, the \nuc{4}{He}-$n$ relative coordinate, unknown
relative wave functions and the total spin of \nuc{5}{He},
respectively. $\Phi_i(^4{\rm He})$ is the configuration of \nuc{4}{He} 
defined in Eq.~(\ref{eq:WF_4He}). Corresponding to the
\nuc{5}{He} wave function, the Hamiltonian of \nuc{5}{He} is given by
\begin{eqnarray}
    H(^{5}{\rm He})
&=& H(^{4}{\rm He}) + \sum_{k=\alpha,n} t_k - T_G(\alpha,n) + V_{\alpha n},
\end{eqnarray}
where $H(^{4}{\rm He})$, $t_k$, $T_G(\alpha,n)$ and $V_{\alpha n}$ are
the internal Hamiltonian of \nuc{4}{He}, the kinetic energy operators
for each cluster (\nuc{4}{He} and $n$), the c.m.motion of the
two-body system, and the \nuc{4}{He}-$n$ interaction, respectively.

The variation of the total energy for the Hamiltonian with respect
to the wave function in Eq.~(\ref{eq:WF_5He}) leads to the
following coupled integro-differential equations:
\begin{eqnarray}
    \sum_{i=1}^6\  \bra \Phi_j(^4{\rm He})| H(^{5}{\rm He}) - E |
    {\cal A}\left\{ [\Phi_i(^4{\rm He}),\chi_i^J(r)]^J \right\}\ket
&=& 0
    \quad
    {\rm for}~j=1,\cdots,6.
    \label{eq:RGM}
\end{eqnarray}

Here we study the behavior of $\{\chi^J_i(r)\}$ and
$\Psi^J(^5{\rm He})$ in the asymptotic region ($r\to\infty$). 
There, the effect of antisymmetrization between
the \nuc{4}{He} cluster and the valence neutron is negligible, and
$V_{\alpha n}$ vanishes. Therefore the coupling between the \nuc{4}{He}
cluster and the valence neutron disappears, and the \nuc{4}{He}
cluster has the same configuration as the isolated \nuc{4}{He}
ground state:
\begin{eqnarray}
    \Psi^J(^5{\rm He})
&~~     \mapleft{r\to\infty}~~&
    \left[ \Psi(^{4}{\rm He}),\chi^J(r) \right]^J,\quad
    {\rm where}\quad
    \Psi(^{4}{\rm He})
~=~ \sum_{i=1}^{6}\ a_i\: \Phi_i
    \label{eq:asympt}.
\end{eqnarray}
 Therefore, it is easy to obtain the following
asymptotic forms of $\{\chi^J_i(r)\}$ from the above 
asymptotic form of $\Psi^J(^5{\rm He})$:
\begin{eqnarray}
    \chi^J_i(r)
&~~     \mapleft{r\to\infty}~~& a_i \cdot \chi^J(r).
    \label{eq:asympt_rel}
\end{eqnarray}
Equation (\ref{eq:asympt_rel}) implies that the asymptotic wave
function $\chi^J_I(r)$ is decomposed into the internal amplitude
$a_i$ of the \nuc{4}{He} cluster and the relative wave function
$\chi^J(r)$ between the \nuc{4}{He} ground state and the valence
neutron. The coefficients $\{a_i\}$ are the same as those in
Eq.~(\ref{eq:WF_4He}), and then the tensor correlation in the
\nuc{4}{He} cluster is the same as that of the isolated \nuc{4}{He}. 
On the other hand, when the valence neutron is close to the
\nuc{4}{He} cluster, the motion of the valence neutron dynamically
couples to configurations of the \nuc{4}{He} cluster in order to
satisfy the Pauli principle. This causes the coefficients $\{a_i\}$ and the
tensor correlation in the \nuc{4}{He} cluster to be different from those 
in the isolated case. 
This implies that $\{a_i\}$ and the tensor correlation depend
on the relative coordinate $r$ via the \nuc{4}{He}-$n$ coupling.

In order to solve the coupled equations of RGM, we apply the
orthogonality condition model (OCM) \cite{Ka99,My02,Sa77}. In the OCM,
antisymmetrization between nucleons in the \nuc{4}{He} cluster and
the valence neutron is replaced by the condition that the relative wave functions $\{\chi_i^J(r)\}$ 
be orthogonal to the Pauli forbidden states due to antisymmetrization. 
Then, integrating over the internal coordinates of the \nuc{4}{He} cluster in the Hamiltonian and
the overlap in Eq.~(\ref{eq:RGM}), we obtain the following coupled
Schr\"odinger equations for the set of relative wave functions $\{\chi_i^J(r)\}$:
\begin{eqnarray}
    \sum_{j=1}^{6} \left[ (T_r + \wtil{V}_{\alpha n} + \Lambda_i)\
    \delta_{ij} + h_{ij}(^4{\rm He}) \right] \chi_j^J(r)
&=& E\ \chi_i^J(r)\quad
    {\rm for}~i=1,\cdots,6,
    \label{eq:OCM}
    \\
    \Lambda_i
~=~ \lambda \sum_{\alpha\in \Phi_i(^{4}{\rm He})} |\phi_\alpha(\wtil{b}_{\alpha})
\ket\bra \phi_\alpha(\wtil{b}_{\alpha})|\quad
&&
    {\rm with}~~~\wtil{b}_{\alpha}
~=~ (\sqrt{5/4})\cdot b_{\alpha}.
\end{eqnarray}
Here $T_r$ and $\wtil{V}_{\alpha n}$ are the kinetic energy
operator for the relative motion and the effective \nuc{4}{He}-$n$
interaction studied in detail below, respectively.
Also, $h_{ij}(^4{\rm He})=\bra \Phi_i | H(^{4}{\rm He}) | \Phi_j \ket$
and $\Lambda_i$ is the projection operator to remove the Pauli
forbidden states $\phi_\alpha(\wtil{b}_{\alpha})$ from the relative wave
functions \cite{Ku86,Ka99}, where $\phi_\alpha(\wtil{b}_{\alpha})$ and
$\alpha$ are the h.o.w.f. with the length parameter $\wtil{b}_{\alpha}$
and the occupied nucleon orbit in each configuration
$\Phi_i(^{4}{\rm He})$, respectively. The value of $\lambda$ is
taken as $10^6$~MeV in the present calculation in order to project out the
components of the Pauli forbidden states into an unphysical energy
region.

Before proceeding to the numerical results, we discuss the modification
of the effective $NN$ interaction for the \nuc{4}{He} cluster in
the extended \nuc{4}{He}+$n$ model. As seen from Eq.~(\ref{eq:OCM}), the \nuc{4}{He}-$n$ interaction
$\wtil{V}_{\alpha n}$ and the internal Hamiltonian matrix $h_{ij}$ of
the \nuc{4}{He} cluster play an important role, in addition to the
antisymmetrization (Pauli) effect described by $\Lambda_i$ in the 
dynamics of couplings between the internal tensor correlation in
the \nuc{4}{He} cluster and the motion of the valence neutron. To
see the deviation in energy of the \nuc{4}{He} cluster due to the
coupling, it is important to reproduce the asymptotic binding
energy of \nuc{4}{He} and the range of the \nuc{4}{He}-$n$
interaction. However, our study presented in the previous section shows
that the tensor correlation tends to shrink the nucleus of \nuc{4}{He}, 
and as a result, the experimental binding energy cannot be reproduced by the employed $NN$ interactions.

\begin{table}[b]
\caption{Properties of \nuc{4}{He} in two effective $NN$
interactions with the observed binding energy (28.3 MeV) and the
matter radius (1.48 fm) being reproduced.} \label{tab:4He3}
\begin{center}
\begin{tabular}{l|cc}\hline\hline
                             &     GA1            &    GA2             \\ \hline
($b_{0s}$,$b_{0p}$)          & (1.385, 0.75)      & (1.385, 0.785)     \\ \hline
($\Delta_v$, $\Delta_R)$     & ($-$0.3234, 0.245) & ($-$0.4179, 0.310) \\ \hline
$\bra V_T \ket$~~  [MeV]~    &  $-$14.5\;~~~         &  $-$29.9\,~~          \\ \hline
$P(2p$-$2h)$~~ [\%]          &   6.9              &   12.5             \\ \hline
$P(\Phi_3)$~~~ [\%]          &   3.8              &   ~\,7.7             \\ \hline
$P(D)$~~     [\%]            &   3.4              &   ~\,6.9             \\ \hline
\end{tabular}
\end{center}
\end{table}

With the above considerations, in order to fit the experimental matter radius and binding energy of
the \nuc{4}{He} cluster, we introduce a phenomenological effective
$NN$ interaction referred to as ``GA1'' by adjusting only the central part, 
while retaining the tensor and LS parts of the effective interaction AK 
considered in the previous section. 
For the central part, we employ GPT, which has a relatively long-range form.
The combination of the GPT central part with the tensor and LS parts of AK was called GA0
in the previous section. However, we know that GA0 gives a radius of \nuc{4}{He} which is small by about 0.1 fm, 
as shown in Table~\ref{tab:4He2}. 
For this reason, in GA1, further modifications needed to reproduce the matter radius are carried out
by changing the second range (attractive part) of the GPT central part as follows:
\begin{eqnarray}
    v_2\cdot e^{-(r/R_2)^2}
&\to&   v'_2\cdot e^{-(r/R'_2)^2}
    ,\quad
    v'_2
~=~     v_2 \cdot (1+\Delta_v),
    \quad
    R_2'
~=~     R_2 + \Delta_R.
\end{eqnarray}
We introduce the two parameters $\Delta_v$ and $\Delta_R$ to reproduce
the binding energy (28.3 MeV) and matter radius (1.48 fm) of
\nuc{4}{He}.  
The results for \nuc{4}{He} using GA1 are presented in Table~\ref{tab:4He3}. 
It is found that GA1 incorporates the tensor correlation, but it cannot reproduce a large amount of the $D$-state
probability. This shortcoming with regard to the $D$-state probability is improved 
by strengthening the tensor force by fifty percent, as discussed in \S\ref{sec:4He}. We call this interaction ``GA2''.

\begin{figure}[t]
\centering
\includegraphics[width=13.0cm,clip]{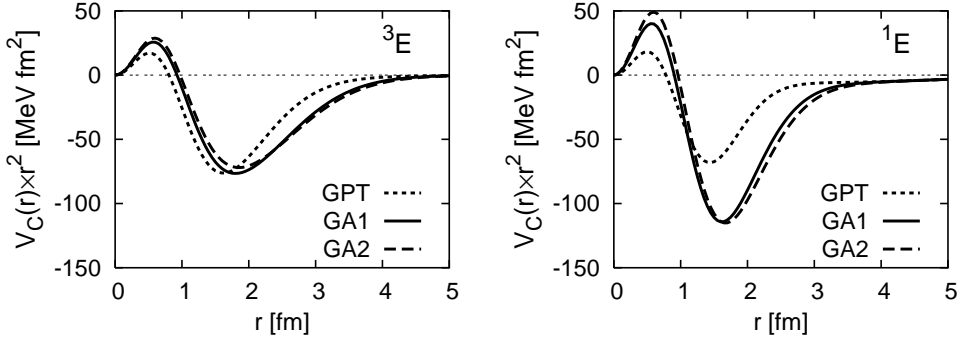}
\caption{The central parts of GA1 (solid curves) and GA2 (dashed curves)
in comparison with that of GPT (dotted curves) for even channels.}
\label{fig:int_C2}
\end{figure}

In Fig.~\ref{fig:int_C2}, we show the form of the central parts of GA1 and GA2 
in comparison to that of GA0, namely, the original GPT central part. 
It is found that the ranges of the central parts of GA1 and GA2 are extended to the outside region 
in order to recover the shrinkage effect of the tensor force. 
This means that the forms of the central forces in the effective interactions with
and without the tensor force must be changed with respect to the range. 
The GA1 and GA2 interactions are regarded as phenomenological effective interactions, 
constructed in order to allow us to analyze the \nuc{5}{He} structures in this study.

\subsection{Effect of the tensor correlation on the doublet splitting and the \nuc{4}{He}-$n$ interaction}

We now solve Eq.~(\ref{eq:OCM}) and determine to what extent the tensor correlation of \nuc{4}{He} 
generates the doublet splitting of $p$-waves in \nuc{5}{He}. For \nuc{4}{He}, we prepare two kinds of
wave functions solved using GA1 and GA2, with different strengths of the tensor correlation. 
For the \nuc{4}{He}-$n$ interaction, $\wtil{V}_{\alpha n}$, in order to elucidate a single effect
of the Pauli blocking from the tensor correlation,  
we consider a Gaussian form for the central part and ignore the LS part, using the following form:
\begin{eqnarray}
    \wtil{V}_{\alpha n}
&=& V_0 \cdot \exp(-(r/R_0)^2)\ .
\end{eqnarray}
The two parameters $V_0$ and $R_0$ are determined so as to reproduce the
energy (0.74 MeV) and the width (0.60 MeV) of the resonant \nuc{5}{He}($3/2^-$) state.\cite{Ao95}
This \nuc{4}{He}-$n$ interaction $\wtil{V}_{\alpha n}$ depends on the wave functions of \nuc{4}{He}
solved with GA1 and GA2 here. 
For the wave functions obtained with GA1, ($V_0$, $R_0$) are ($-$62.60 MeV, 2.15 fm), 
and for those obtained with GA2, ($V_0$, $R_0$) are ($-$77.96 MeV, 1.95 fm).

The results for two $p$-wave resonant poles are listed in
Table~\ref{tab:5He}.  It is found that both cases produce visible
splits in the energies of the $3/2^-$ and $1/2^-$ states without the LS force, 
and that GA2 gives a value that is twice as large as that given by GA1,
due to the strong Pauli-blocking effect. For comparison, we also
calculate poles with the so-called KKNN \nuc{4}{He}-$n$ potential\cite{Ka79} 
with a simple \nuc{4}{He}[$(0s)^4$]+$n$ model. 
The KKNN potential is a semi-phenomenological \nuc{4}{He}-$n$ interaction
possessing central and LS parts and is constructed to fit the observed
$s$- and $p$-wave phase shifts based on the RGM analysis, 
assuming the $(0s)^4$ configuration of \nuc{4}{He}.
Therefore, there is no tensor correlation of \nuc{4}{He} and 
no Pauli blocking for $p$-waves in the KKNN potential, and 
the observed splitting is accounted for by only the LS part. In comparison to the KKNN
potential, GA2 produces about half of the splitting energy.
This is consistent with the results of a previous study\cite{Na59}.

\begin{table}[t]
\caption{The positions of the $p$-wave resonant poles (energy and width) of 
\nuc{5}{He}(3/2$^-$,1/2$^-$) measured from the
\nuc{4}{He}+$n$ threshold energy for three interactions. $\Delta
E$ is the splitting energy. All values are in units of MeV.}
\label{tab:5He}
\begin{center}
\begin{tabular}{c|cccc}\hline\hline
            &  GA1          &  GA2          & KKNN         \\\hline
3/2$^-$     &  (0.74,~0.60) &  (0.74,~0.60) & (0.74,~0.60) \\
1/2$^-$     &  (1.10,~1.45) &  (1.47,~3.10) & (2.13,~5.84) \\\hline
$\Delta E$  &    0.36       &    0.73       &     1.39     \\\hline
\end{tabular}
\end{center}
\end{table}
\begin{figure}[t]
\centering
\includegraphics[width=13.0cm,clip]{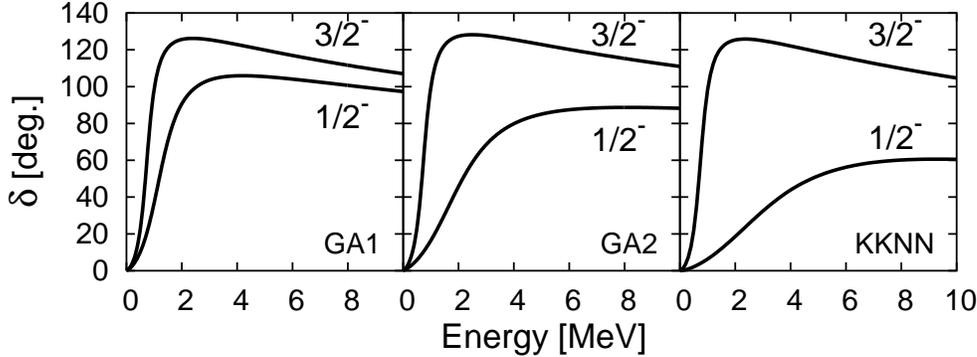}
\caption{$p$-wave phase shifts of the \nuc{4}{He}+$n$ system
measured from the \nuc{4}{He}+$n$ threshold energy for three interactions.}
\label{fig:5He_ph1}
\end{figure}

In Fig.~\ref{fig:5He_ph1}, we show the $p$-wave phase shifts of
the extended \nuc{4}{He}+$n$ model using three interactions
corresponding to Table~\ref{tab:5He}. We used the level density
formalism in the complex scaling method to calculate these phase shifts
\cite{Sz04}. 
It is seen that the splittings in the phase shifts are similar to cases of resonant poles. 
In the results, the mechanism of the doublet splitting is understood from the viewpoint of the
tensor correlation via the Pauli blocking. Our results also
indicate that the conventional LS interaction determined to
reproduce the observed splitting energy should be weakened when
the tensor correlation is taken into account explicitly.

It is noted that in some previous studies\cite{Te60}, 
it is claimed that the origin of the doublet splitting in \nuc{5}{He} is the difference
between the multiplicities of the $0p_{1/2}$ (two) and $0p_{3/2}$(four) orbits,
which lead to different strengths of the Pauli-blocking. 
However, in our study, this effect contributes only 20\%-30\%  of the splitting, and a large mixing of the
$0p_{1/2}$ orbit in \nuc{4}{He} caused by the tensor force has a much more important effect. 
The latter is interpreted as a pion effect considered in other studies\cite{To02,Su04,Og04}.

Based on the knowledge obtained in the above analysis, we now
construct a reliable \nuc{4}{He}-$n$ interaction $\wtil{V}_{\alpha
n}$ including the tensor correlation in the \nuc{4}{He} cluster,
which reproduces the observed phase shifts of this system. This
interaction is useful for the quantitative studies of the \nuc{5}{He} structure.

\begin{table}[t]
\caption{Values of the four parameters in the KKNN-T1 and KKNN-T2 potentials}
\label{tab:KKNN-T}
\begin{center}
\begin{tabular}{c|cccc}\hline\hline
         & $a_C$        & $a^P_C$     & $a_{LS}$    & $V^C_{2}$ [MeV] \\\hline
KKNN-T1  & 1.20         &  0.78       &  0.70       &  100.0      \\
KKNN-T2  & 1.36         &  0.67       &  0.52       &  120.0      \\\hline
\end{tabular}
\end{center}
\end{table}

We start from the KKNN potential which consists of the central
and LS parts in the following:
\begin{eqnarray}
    \wtil{V}^{\mbox{\scriptsize KKNN}}_{\alpha n}
&=& V_C(r) + V^P_C(r)\cdot P_r + V_{LS}(r) + V^P_{LS}(r)\cdot P_r,
    \label{eq:KKNN}
\end{eqnarray}
where $P_r$ is the parity operator with respect to the relative
coordinate $r$.  The radial forms of each term in Eq.~(\ref{eq:KKNN})
are described by a linear combination of two- or three-range
Gaussians\cite{Ka79}. Next, we attempt to modify the KKNN potential 
in such a way that it cooperates with the blocking effect of the tensor correlation in
the \nuc{4}{He}+$n$ system. As discussed in the previous
subsection, the blocking effect can be quantitatively represented by
repulsive potentials depending on the $p$-wave doublet states,
and the resultant difference between the resonance energies of 
the doublet states can be understood as the effect of an LS
interaction. Accordingly, we constructed a new \nuc{4}{He}-$n$
interaction by modifying the form of the KKNN potential as follows:
\begin{eqnarray}
    \wtil{V}^{\mbox{\scriptsize KKNN-T}}_{\alpha n}
&=& a_C\cdot V'_C(r) + a^P_C\cdot V^P_C(r)\cdot P_r
 +  a_{LS}\cdot \bigl\{ V_{LS}(r) + V^P_{LS}(r)\cdot P_r\bigr\}.~
\end{eqnarray}
This potential, which we call the KKNN-T potential, has four
parameters, $a_C$, $a^P_C$, $a_{LS}$ and the potential strength of
the repulsive part in $V'_C(r)$, which is expressed by $V^C_2$.
All the other parts of the $\wtil{V}^{\mbox{\scriptsize KKNN-T}}_{\alpha n}$
are taken to be unchanged from the original KKNN potential. We chose the
set of the four parameters ($a_C$,~$a^P_C$,~$a_{LS}$,~$V^C_2$) so
as to reproduce the observed $s$- and $p$-wave phase shifts by
solving the OCM Eq.~(\ref{eq:OCM}) in the extended \nuc{4}{He}+$n$
model. We construct potentials with two parameter sets, called KKNN-T1 and -T2, 
incorporating the tensor correlation induced by GA1 and GA2, respectively. 
The parameter sets of the KKNN-T1 and -T2 potentials are listed in Table \ref{tab:KKNN-T}.

\begin{figure}[t]
\centering
\includegraphics[width=12.7cm,clip]{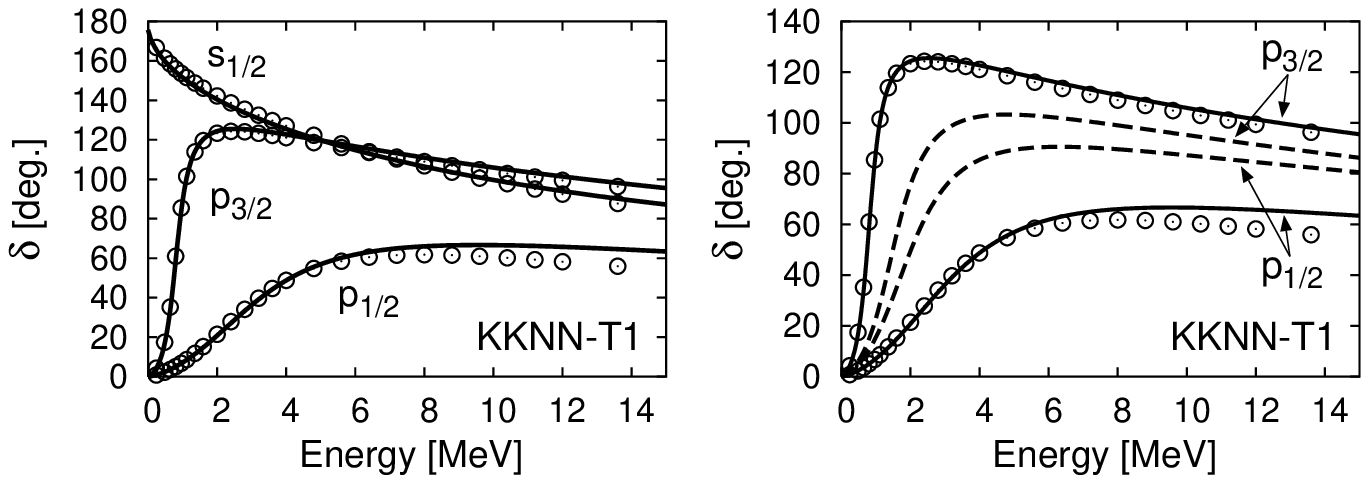}
\includegraphics[width=12.7cm,clip]{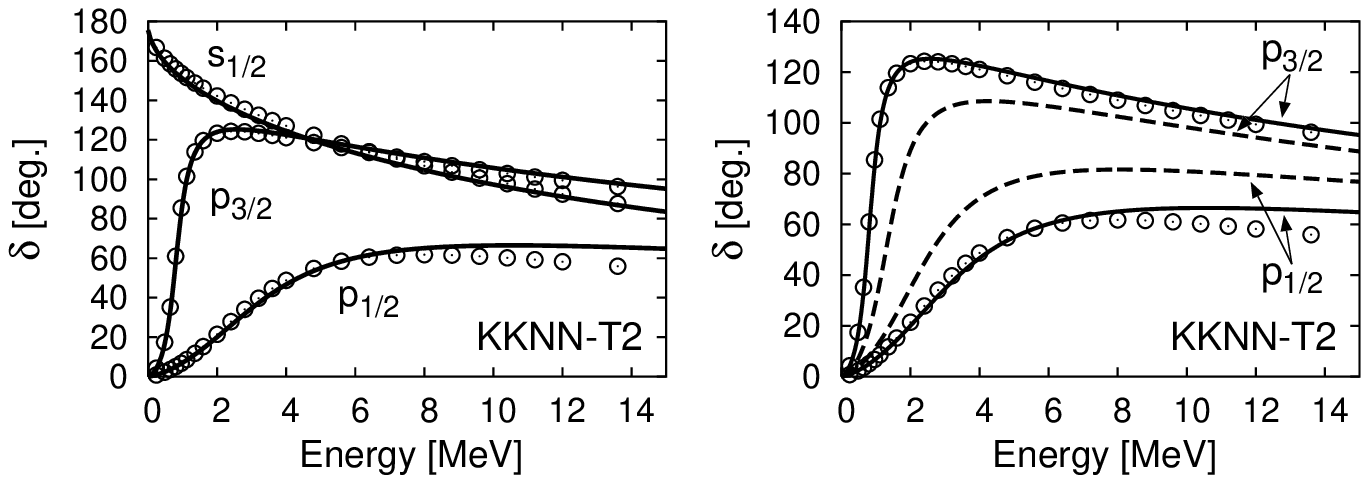}
\caption{Left: $s$- and $p$-wave phase shifts for the \nuc{4}{He}+$n$
system with KKNN-T1 (upper) and KKNN-T2 (lower) in comparison with 
the experimental results\cite{St72} represented by the circles. 
Right: Decomposition of the $p$-wave doublet splitting. The dashed curves represent the calculations
without the LS force in the KKNN-T1 (upper) and KKNN-T2 (lower) potentials.}
\label{fig:5He_ph2}
\end{figure}

Since we know that the tensor correlation contributes to the
$p$-wave doublet splitting, the strength of the LS part must be
weaker than that in the original KKNN potential, as seen from the values
of $a_{LS}$ in Table~\ref{tab:KKNN-T}. This reduction effect is
stronger for the KKNN-T2 potential ($a_{LS}=0.52$) than for the KKNN-T1 potential ($a_{LS}=0.70$). 
We find that the central part for the $p$-waves is
slightly strengthened in total, and the weak Pauli-blocking
effect and the weak contribution from the LS part for the
$p_{3/2}$ state of \nuc{5}{He} are recovered. The obtained $s$- and $p$-wave phase
shifts are shown in Fig.~\ref{fig:5He_ph2}. We see good agreement
with the experimental data\cite{St72} in both cases. 
In this study, the splitting of $p$-wave phase shifts is decomposed into 
two kinds of contributions that from the Pauli blocking 
due to the tensor correlation and that from the LS part 
in the \nuc{4}{He}-$n$ interaction. 
In the right panels of Fig.~\ref{fig:5He_ph2}, we can divide 
the two contributions using the results obtained by omitting the LS part in KKNN-T1 and -T2 potentials. 
It is then found that the KKNN-T2 potential gives a large splitting 
from the tensor correlation, about  half of the observed one.

\begin{figure}[t]
\centering
\includegraphics[width=12.7cm,clip]{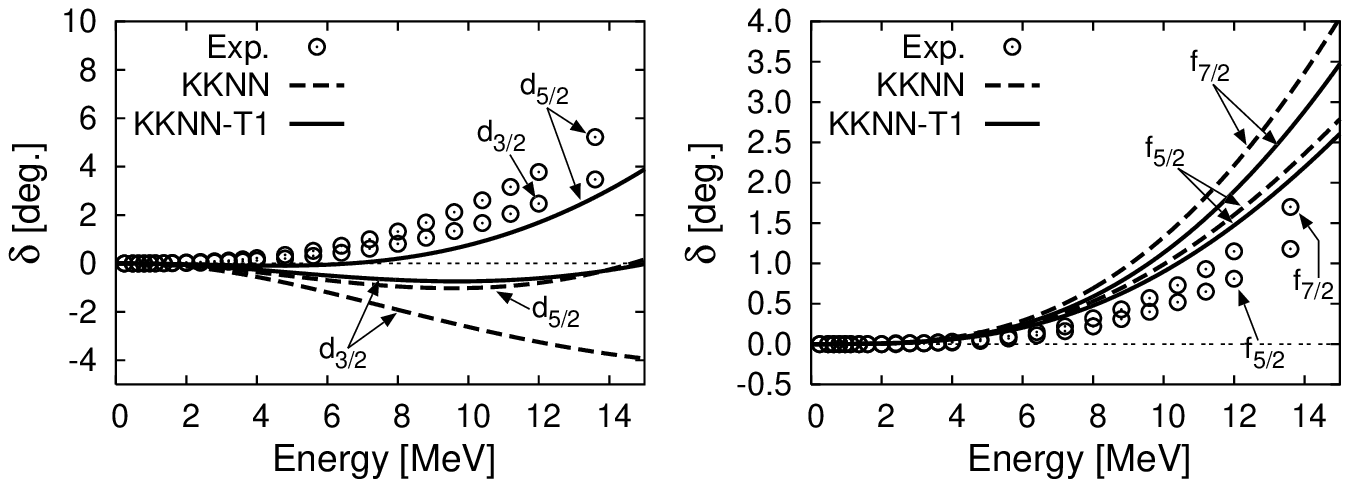}
\includegraphics[width=12.7cm,clip]{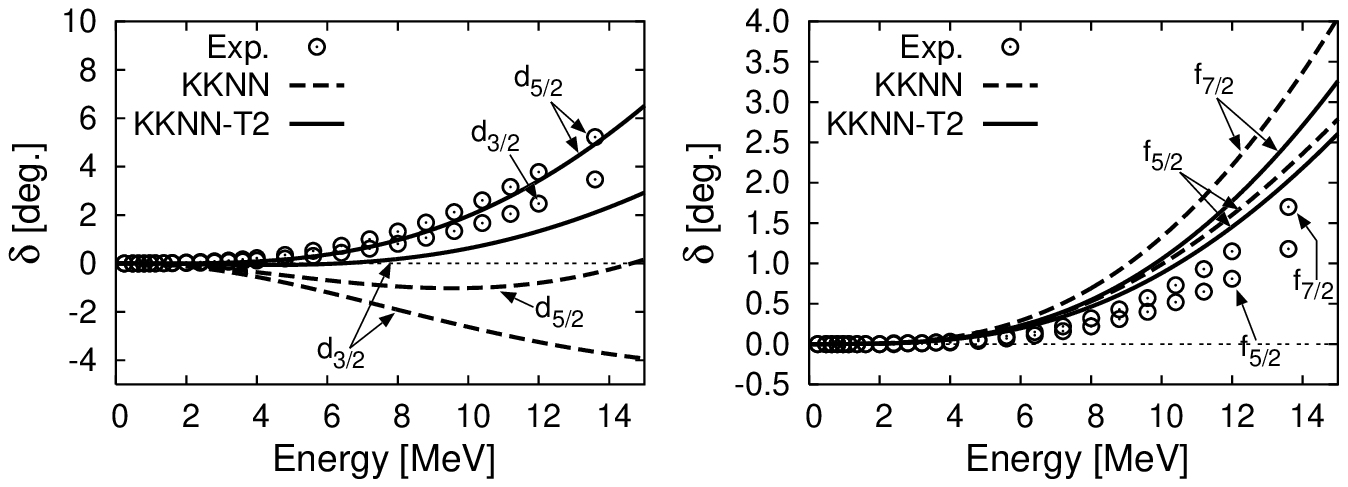}
\caption{$d$- and $f$-wave phase shifts of the \nuc{4}{He}+$n$ system
obtained with the KKNN-T1 (upper) and KKNN-T2 (lower) potentials in comparison to those obtained with the 
KKNN potential and the experimental results\cite{St72}.
}
\label{fig:5He_ph3}
\end{figure}

In order to show the reliability of the KKNN-T1 and -T2 potentials, 
we also investigate the $d$- and $f$-wave phase shifts
and compare the results obtained using these potentials to those obtained using the KKNN potential shown in
Fig.~\ref{fig:5He_ph3}. The KKNN potential cannot reproduce the
experimental data\cite{Ao95,St72}, in particular, the $d$-waves exhibit
repulsive behaviors, and doublet splittings are larger than the 
experimental values in both cases. 
Contrastingly, for the KKNN-T1 and -T2 potentials 
with the extended \nuc{4}{He}+$n$ model, these discrepancies are decreased,
and calculated phase shifts are closer to their experimental values. 
The improvement is confirmed in the KKNN-T2 potential. This success results from the fact 
that the tensor correlation in \nuc{4}{He} does not contribute to the $d$- and $f$-waves 
as much as in the case of the KKNN potential and that the weakening of the LS part in the
KKNN-T1 and -T2 potentials works well to reproduce the observed
small splittings. These results imply that the description of
higher partial waves in \nuc{5}{He} are naturally improved by
considering the tensor correlation.

The ratio of the two contributions of the tensor correlation and 
the LS force in the doublet splitting depends on the strength of
the tensor correlation in the \nuc{4}{He} cluster. This point will
be investigated quantitatively in a subsequent study using a hybrid
$V$+$T$ model of \nuc{4}{He}, in order to make the evaluation of the tensor
correlation firmer.

The mechanism of the doublet splitting in \nuc{5}{He} is related
to the structure of \nuc{6}{He} concerning the $0p_{1/2}$ orbit,
in particular, resonances\cite{My01}. To examine this, it is
interesting to carry out an extended \nuc{4}{He}+$n$+$n$ model
analysis by considering the tensor correlation in \nuc{4}{He}.

\section{Summary}\label{sec:summary}

We have investigated the role of the tensor correlation 
in the structure of \nuc{4}{He} and its effect on \nuc{5}{He}. 
From the meson theory with a rigorous treatment of the NN interaction,
the contribution of the tensor force to the binding energy of \nuc{4}{He}
is found to be comparable to that of the central force. 
But in most nuclear models, the tensor force is not treated
explicitly, and the physical properties of the tensor correlation are not clarified. 
The purpose of this paper is to treat the tensor force explicitly
and to understand the essential effects of the tensor correlation.
For this purpose, \nuc{4}{He} and \nuc{5}{He} are good objects of studies, due to the simpleness of their structures.

In our study, we adopt a similar framework to the CPPHF proposed by Sugimoto et al.\cite{Su04}, 
but a simpler and more conventional shell model approach.
For \nuc{4}{He}, we have extended the shell model type wave function from the
original $(0s)^4$ configuration to the $(0s)^4$+$(0s)^2(0p)^2$ configurations,
taking into consideration the properties of the tensor force; that is, in the wave function,
all the $2p$-$2h$ configurations of $(0s)^2(0p)^2$ can be coupled with
the original $(0s)^4$ configuration by the tensor force.  
The length parameters of the $0p$ orbits were chosen to be free in order to 
take into account the high momentum components induced by the tensor force.
It is found that the obtained wave function composed of the superposition
of all the configurations with spatially modified $0p$ orbits describes 
the basic characteristics of the tensor correlation, and that they agree with
those obtained in CPPHF.\cite{Su04}
The numerical results show that the tensor correlation has two characteristics:
One is a strong coupling of the $0s_{1/2}$ orbit to the $0p_{1/2}$ orbit, which 
reflects the $0^-$ coupling of the pion nature,
and the other is a large mixing of the excited two-nucleon pair of 
the $(0p_{1/2})^2$ configuration, which shows a similar correlation in deuteron.

For \nuc{5}{He}, we have studied the Pauli-blocking effect of the tensor correlation 
on the $p$-wave doublet splitting. 
We clarified that because the $0p_{1/2}$ orbit is preferentially mixed in the $2p$-$2h$ components of \nuc{4}{He} 
induced by the coupling from the tensor force with the pion nature, 
there arises a coupling produced dominatingly by the Pauli principle 
between the $2p$-$2h$ configurations of \nuc{4}{He} and the $p_{1/2}$ valence neutron in \nuc{5}{He}.
This coupling was treated by solving the coupled OCM equation in the extended \nuc{4}{He}+$n$ model.
The calculated results show that the tensor correlation causes about 
half of the $p$-wave doublet splitting in \nuc{5}{He},
and the pion nature of the tensor correlation plays an important role in the splitting.
This finding cannot be accounted for within the old understanding\cite{Te60,Na59}.
Our explicit treatment of the tensor correlation leads to a weakening of 
the conventional strength of the LS force. 
In fact, we have a newly constructed \nuc{4}{He}-$n$ interaction realized through the reduction of 
the strength of the LS part by introducing the tensor correlation in \nuc{4}{He}. 
This new interaction not only reproduces the $s$- and $p$-wave properties 
but also improves the $d$- and $f$-wave phase shifts. 
These results demonstrate that the tensor correlation is necessary to reproduce 
all the phase shifts of the neutron scattering with \nuc{4}{He} in various partial waves.

The tensor correlation depends strongly on the coupling matrix element of the tensor force.
Therefore, one important problem is to study the coupling from the tensor force, as discussed in \S\ref{sec:4He}.
We would like to extend our model wave function for \nuc{4}{He} to the hybrid $V+T$ bases in future analyses.

\section*{Acknowledgements}

The authors would like to thank Prof. H. Toki for fruitful discussions and encouragement.
We also would like to acknowledge valuable discussions with Dr. S. Sugimoto. 
We thank Prof. Akaishi for providing us the results of 
the $G$-matrix calculation for the AV8' potential. 
One of the authors (T. Myo) thanks Mr. R. Suzuki 
for helping with the calculation of the phase shifts of the \nuc{4}{He}+$n$ system
using level densities calculated with the complex scaling method.
This work was performed as a part of the ``Research Project for
Study of Unstable Nuclei from Nuclear Cluster Aspects (SUNNCA)''
sponsored by RIKEN.

\end{document}